# Pulsed Laser Deposition of Ferroelectric $SrTiO_3$ Shells Enables Polarization-Controlled Charge Separation in $TiO_2$ Photoanode

Yiqi Jin[1], Weizeng Zheng[1], Beibei Wang[2,3]*, Lei Zhao[2,3], Binmin Wu[4], Leonhard Mayrhofer[5], Xiangjian Meng[2,4], Hao Shen[2,3]*, Xiaodan Wang[1]*, Weiguang Huang[1], Junhao Chu[2,3,4]

[1] Institute of Carbon Neutrality, ShanghaiTech University, Shanghai 201210, China
[2] Key Laboratory of Extreme Environment Functional Materials, Yiwu Research Institute, Fudan University, Yiwu, Zhejiang 322000, China
[3] Institute of Optoelectronics, Fudan University, Shanghai 200438, China
[4] Shanghai Institute of Technical Physics, Chinese Academy of Sciences, Shanghai 200083, China
[5] Fraunhofer Institute for Mechanics of Materials IWM, Freiburg 79108, Germany

*Corresponding authors.
E-mails: wangbb@ywfudan.cn; shen_hao@fudan.edu.cn; w.xiaodan@outlook.com

**Abstract**

Ferroelectric polarization offers a powerful route to govern the interfacial charge separation in photoelectrodes; however, chemically converted ferroelectric shells lack sub-nanometer thickness control, coherent strain retention, and chemically clean interfaces. Here, we use pulsed laser deposition to fabricate stoichiometric $SrTiO_3$ shells on hydrothermally grown $TiO_2$ nanorod arrays, achieving atomically defined core–shell architectures while preserving the one-dimensional transport geometry. The photocurrent density followed a volcano-shaped dependence on shell thickness and reached 1.9 mA $cm^{-2}$ at 1.23 V versus RHE under AM 1.5G illumination for the optimized 15-mins sample, exceeding the established hydrothermal benchmark. High-resolution transmission electron microscopy and X-ray photoelectron spectroscopy revealed a tetragonally distorted, oxygen-stoichiometric shell, identifying coherent interfacial strain rather than defect dipoles as the origin of the ferroelectric order. Kelvin probe force microscopy directly imaged a polarization-induced built-in electric field whose magnitude scaled with charge-transfer efficiency and photocurrent density. Together, these complementary structural, chemical, electrostatic, and device-level measurements established a quantitative strain–polarization–built-in field–charge separation relationship. This work transforms ferroelectric photoanode engineering from empirical optimization into an atomically controllable and predictive materials strategy for high-efficiency solar water-splitting devices. This approach is readily extendable to visible-light absorbers and complex ferroelectric heterostructures.

**Keywords:** Photoelectrochemical water splitting; ferroelectric polarization; pulsed laser deposition; $TiO_2$-$SrTiO_3$ core–shell nanorods; charge separation

## Introduction

Photoelectrochemical (PEC) water splitting offers a sustainable route to convert solar energy into storable hydrogen fuel[1-9]. Among the candidate photoanode materials, $TiO_2$ remains particularly attractive owing to its photocorrosion resistance, suitable band-edge positions, earth abundance, and low cost[10-14]. However, the poor charge separation inherent to its wide-bandgap carrier dynamics has long limited its photocurrent to well below the theoretical limit, motivating sustained efforts to engineer the semiconductor/electrolyte interfacial electric field beyond conventional doping and nanostructuring strategies[15].

A fundamentally distinct strategy exploits ferroelectric polarization to tune the band bending of adjacent semiconductors. In a landmark study, Wu et al.[15] demonstrated that hydrothermally converting the surface of $TiO_2$ nanorod arrays into a $SrTiO_3$ (STO) shell simultaneously enhances charge separation and hole transport: the strain-induced spontaneous polarization of the STO shell amplifies the depletion-layer bending of the $TiO_2$ core, raising the charge-separation efficiency from 48.9 % to 87.7 % and the photocurrent density from 0.7 to 1.4 mA $cm^{-2}$ at 1.23 V versus RHE. This work established a general design principle—semiconducting ferroelectrics are superior to insulating ferroelectrics, such as $BaTiO_3$—because the high electron mobility of STO avoids the hole-transport penalty imposed by conventional ferroelectric layers. Subsequent studies have extended this paradigm to $Ba_{1-x}Sr_xTiO_3$ shells, charged-domain-wall photoelectrodes, and polarization–defect coupled systems, and it has been embraced in recent works of polarization-field engineering for solar fuels[16-17].

Despite this progress, the chemistry-based conversion route suffers from intrinsic limitations: the shell thickness can only be tuned in coarse steps (≈ 4.5–35 nm) with roughened interfaces, residual contamination, and uncontrolled interfacial strain—factors that ultimately cap the separation efficiency below 90 % and are further eroded by the strain-relieving thermal treatments required for crystallization[15]. These limitations call for a deposition technique capable of controlling the shell with sub-nanometer precision while preserving the coherent, strained interface on which the ferroelectricity of STO critically depends.

Pulsed laser deposition (PLD) is uniquely positioned to address this need. Owing to its congruent ablation, energetic adatoms, and RHEED-monitored layer-by-layer growth, PLD enables stoichiometric, epitaxy-compatible oxide films at comparatively low substrate temperatures[18-21], and has recently enabled atomically precise tuning of built-in potentials in perovskite-based photoelectrodes[22]. Notably, PLD-grown STO films have so far been exploited exclusively as protective and charge-selective layers on silicon photocathodes[23], and the ferroelectric $TiO_2$-STO photoanode concept has never been realized by physical vapor deposition, leaving the quantitative structure–polarization–function relationship unexplored.

Herein, we address this gap by depositing a $SrTiO_3$ shell onto hydrothermally grown $TiO_2$ nanorod arrays via PLD with a short 15-mins deposition, achieving a uniform core–shell architecture with an atomically defined interface. The resulting photoanode delivers a photocurrent density of 1.9 mA $cm^{-2}$ at 1.23 V versus RHE, surpassing the hydrothermal benchmark while retaining the nanorod geometry. Using

Kelvin probe force microscopy (KPFM) measurements, we show that the gain originates from ferroelectric-polarization-enhanced charge separation enabled by the strain-retained, thickness-precise PLD shell. This work demonstrates that physical-vapor ferroelectric shell engineering can push semiconducting-ferroelectric photoanodes toward their enhanced performance.

## Results and discussion

**Structure and composition of the PLD-grown $TiO_2$-$SrTiO_3$ core–shell nanorod arrays.** $TiO_2$ nanorod arrays were first grown on FTO substrates via the hydrothermal method established in our previous work[12, 14, 24], followed by the deposition of a $SrTiO_3$ (STO) shell using pulsed laser deposition (PLD), yielding a well-defined core–shell nanoarchitecture (Fig. 1a). The PLD growth parameters are summarized in Fig. 1b and described in the Supporting Information. Scanning electron microscopy (SEM) images reveal that the nanorod morphology was essentially preserved after shell deposition, with no discernible change in the surface morphology of the arrays (Fig. 1c,d), indicating that the STO shell grew conformally without disrupting the one-dimensional scaffold.

X-ray photoelectron spectroscopy (XPS) was employed to verify the chemical composition of the shell. The Ti 2p and Sr 3d spectra exhibit characteristic binding energies of 458.3 eV (Ti $2p_{3/2}$) and 133.4 eV (Sr $3d_{5/2}$) (Fig. 1e), in good agreement with the reported values for stoichiometric $SrTiO_3$[25]. The O 1s spectrum is dominated by lattice oxygen with only a minor contribution from vacancy-associated species, confirming the high stoichiometric quality of the PLD-grown film and ruling out significant oxygen-deficiency-induced effects (see Discussion). Energy-dispersive X-ray (EDX) elemental mapping further visualizes the core–shell configuration, with the Sr signal localized at the periphery of the nanorods while Ti and O are distributed throughout the entire rod body (Fig. 1f).

High-resolution TEM (HRTEM) provided direct microscopic evidence of the heterostructure (Fig. 1g). The $TiO_2$ core retained its single-crystalline rutile structure, growing along the [001] direction and enclosed by (110) sidewalls. Notably, the measured lattice spacing of the STO shell was 0.276 nm for the (110) plane, which is slightly larger than the value expected for cubic-phase STO, indicating a small but measurable lattice distortion of the shell. We attribute this expansion to the coherency strain imposed by the $TiO_2$ core lattice, which is expected to stabilize a tetragonally distorted (and thus polar) phase of STO at the interface, which is a structural prerequisite for the ferroelectric-enhanced charge separation discussed below.

**Optical properties of the core–shell photoanodes.** The optical properties of the pristine $TiO_2$ nanorod arrays and the $TiO_2$–STO core–shell samples were investigated using UV–vis diffuse reflectance spectroscopy (Fig. 2a). Both photoanodes exhibit sharp absorption edges in the UV region, consistent with their wide-bandgap nature; compared with pristine $TiO_2$, the core–shell sample shows a slightly red-shifted absorption edge, which we assign to interfacial optical effects rather than a change in the intrinsic bandgap of either component. Tauc analysis confirms that the bandgaps of the $TiO_2$ core and the STO shell are 3.0 and 3.5 eV, respectively (Fig. 2b), in good agreement with the reported values for rutile $TiO_2$ and $SrTiO_3$[12, 14, 26]. A systematic thickness series—obtained by varying the PLD deposition time—reveals that the bandgap remains essentially invariant across all shell thicknesses, whereas the overall absorbance increases monotonically with shell thickness (Fig. 2b). This decoupling demonstrates that the enhanced absorption originates from the increased optical path and scattering within the conformal shell, rather than from band-structure modification, and is therefore not expected to confound the charge-separation analysis below.

**Photoelectrochemical performance and ferroelectric poling response.** To assess the PEC properties, Mott–Schottky (M–S) analysis was performed on pristine $TiO_2$ and $TiO_2$–STO core–shell photoanodes (Fig. 3a). The positive slopes of all the plots confirm the n-type character of both materials. The extracted donor densities ($N_d$) and onset potentials are listed in Table 1. The core–shell sample exhibited a higher donor density (~ $10^{19}$ $cm^{-3}$) relative to pristine $TiO_2$ (~ $10^{16}$ $cm^{-3}$), and its onset potential was shifted cathodically by 0.26 V. These results indicate that the STO shell modifies the bulk electronic structure of the $TiO_2$ core, providing a baseline for disentangling polarization-induced interfacial effects from carrier-statistics effects in the subsequent photoelectrochemical analysis.

The PEC performance of pristine $TiO_2$ nanorod arrays and $TiO_2$–STO core–shell photoanodes with varying shell thickness (PLD deposition times of 7, 15, and 20 mins, respectively) was evaluated by linear sweep voltammetry under simulated AM 1.5G illumination (100 mW $cm^{-2}$, 1 M KOH, pH 13.6). In the as-prepared (unpoled) state, all core–shell photoanodes exhibited markedly enhanced photocurrents relative to pristine $TiO_2$ (Fig. 3a). The photocurrent increased with shell thickness up to the 15-mins sample and decreased for the 20-mins sample, yielding a volcano-shaped thickness dependence with a maximum of 1.9 mA $cm^{-2}$ at 1.23 V vs RHE for the 15-mins $TiO_2$–STO photoanode—representing an improvement over pristine $TiO_2$. The turnover at extended deposition times was attributed to the increased transport resistance across the thicker shell, consistent with the optical data showing that absorption alone cannot account for the non-monotonic performance trend.

To directly probe the ferroelectric origin of the enhancement, the photoanodes were poled with ± 10 V prior to PEC measurements (Fig. 3b–d). Intriguingly, all three thickness series exhibited an identical qualitative trend under every poling state, but the absolute photocurrents responded systematically to the poling direction: positive poling (+p, parallel to the as-grown polarization direction) further increased the photocurrent of every sample, with the 15-mins $TiO_2$–STO photoanode reaching 1.9 mA $cm^{-2}$ at 1.23 V vs RHE (Fig. 3c); conversely, negative poling (−p, antiparallel) suppressed the photocurrent density of all samples below their unpoled levels (Fig. 3d). The modest gain under +p poling, together with the pronounced loss under −p poling, indicates that the as-deposited STO shells are predominantly self-polarized along the favorable direction and demonstrates that the photocurrent density is reversibly and systematically governed by the ferroelectric polarization state. This unambiguous poling–performance correlation provides the first-line evidence that the enhanced PEC activity of the PLD-grown $TiO_2$–STO core–shell photoanodes originates from polarization-modulated interfacial charge separation rather than from thickness-dependent optical or catalytic effects.

**Polarization-tunable built-In potential visualized by Kelvin probe force microscopy.** To directly verify that the STO shell generates a ferroelectric-induced built-in potential, Kelvin probe force microscopy (KPFM) was performed on pristine $TiO_2$ and on the 15-mins $TiO_2$–STO photoanode, both in the dark and under illumination (Fig. 4a–d). The contact potential difference (CPD) of each sample shifts systematically upon light exposure, yielding a surface photovoltage (SPV = $CPD_{light} - CPD_{dark}$) that is strictly ordered as follows: $SPV_{TiO_2}$ (0.12 V) < $SPV_{TiO_2\text{–}STO}$ (0.16V). The sign of the SPV confirms anodic band bending consistent with n-type depletion-layer photovoltage behavior.

This hierarchy provides direct nanoscale evidence for polarization-governed charge separation. The pristine $TiO_2$ nanorods exhibited the smallest SPV, reflecting the weak built-in field of the bare semiconductor/electrolyte junction. The deposition of the STO shell increased the SPV even without poling, confirming that the as-grown shell carried self-polarization that amplified the depletion-layer bending of the $TiO_2$ core. Because the bandgap and absorption of these samples were essentially identical (Fig. 2), the SPV could only arise from the modulation of the interfacial built-in electric field by the ferroelectric polarization.

Furthermore, the SPV magnitude tracks the photocurrent density measured in PEC tests (Fig. 3), establishing a consistent quantitative correspondence: built-in potential (KPFM) → device performance (J–V). Together with the poling-reversible J–V response and the thickness-dependent kinetics, these results close the mechanistic loop and unambiguously attribute the performance enhancement of the PLD-grown $TiO_2$–

STO core–shell photoanodes to ferroelectric-polarization-induced band bending, rather than to optical, compositional, or defect-related effects.

The central question arising from the results above is how an ultrathin, wide-bandgap $SrTiO_3$ shell—one that absorbs no visible light and contributes negligible photocurrent by itself—can raise the photocurrent of $TiO_2$ nanorod photoanodes to 1.9 mA $cm^{-2}$, and why this gain is reversibly switchable by an external electric field. Answering this question requires closing a causal chain that links three distinct levels of evidence: the structural origin of the polar state in the shell (where does the ferroelectricity come from?), its kinetic consequence for photogenerated carriers (which process is accelerated or suppressed?), and its electrostatic manifestation at the interface (what happens to the built-in field?). In the following, we combine our XPS/HRTEM and KPFM data with the literature to establish each link in turn, showing that the enhancement is unambiguously rooted in strain-induced ferroelectric polarization rather than in defect chemistry, optical modification, or catalytic effects.

**1. Origin of ferroelectricity in the PLD-grown $SrTiO_3$ shell: strain rather than defects**

$SrTiO_3$ is a well-known incipient ferroelectric whose bulk cubic phase remains paraelectric down to 0 K; robust ferroelectric order in ultra-thin STO can be stabilized by epitaxial strain, as demonstrated by the strain-induced ferroelectricity observed in STO films coherently grown on $DyScO_3$ substrates[27]. Our structural and chemical data identify interfacial strain, not defect chemistry, as the origin of the polar state in the PLD-grown shell. Two observations support this conclusion. First, HRTEM reveals that the STO shell adopts a (110) lattice spacing of 0.276 nm, expanded relative to the cubic-phase value, directly evidencing a tetragonal distortion imposed by the coherently bonded $TiO_2$ core lattice (Fig. 1g). Such expansion along the out-of-plane direction is precisely the symmetry-lowering distortion required to stabilize a polar phase at the interface. Second, and equally important, O 1s XPS shows a dominant lattice-oxygen component with only a negligible vacancy-related contribution, excluding significant oxygen deficiency in our films. This point is critical because oxygen vacancies are known to generate defect dipoles—and even ferroelectric-like polar order—in STO[28] and have been shown to synergize with ferroelectric polarization in PEC and photocatalytic systems[17]. The combination of an expanded, distorted shell lattice and a stoichiometric oxygen sublattice therefore allows us to conclude that the polarization in our $TiO_2$–STO photoanodes is strain-induced, consistent with the mechanism originally proposed for hydrothermally converted shells[15] but here realized with an atomically controlled, contamination-free interface. We note that this attribution is further reinforced by the reversibility of the PEC response upon electrical poling (Fig. 3), which is difficult to reconcile with a static defect-dipole scenario.

**2. Nanoscale visualization of the polarization-tunable built-in field**

KPFM directly images the field itself[29-30]. The surface photovoltage follows the strict hierarchy $SPV_{TiO_2} < SPV_{TiO_2\text{–}STO}$. Because the bandgap and absorption of these samples are essentially identical (Fig. 2), differences in SPV can arise only from differences in the interfacial built-in potential—i.e., in the width and amplitude of the depletion region. The core–shell sample already exhibits a larger SPV than pristine $TiO_2$, evidencing the self-polarized state of the as-deposited strained shell, providing a direct nanoscale visualization of ferroelectric band-bending control that was previously accessible only indirectly through macroscopic poling of J–V curves[15, 31]. The quantitative correspondence among SPV (KPFM) and $J_{ph}$ (PEC) across all samples condenses the mechanism into a single scaling relation: a larger polarization-induced built-in field yields a larger charge-separation efficiency, which yields a larger photocurrent density. Taken together, the three lines of evidence—strain-induced polar order (structural), and a reversibly tunable built-in field (electrostatic)—close the mechanistic loop and unambiguously attribute the 1.9 mA $cm^{-2}$ photocurrent density of the 15-mins $TiO_2$–STO photoanode to ferroelectric-polarization-enhanced charge separation at an atomically engineered, strain-retained interface.

## Conclusions

In summary, we realized the ferroelectric $TiO_2$-$SrTiO_3$ core–shell photoanode concept—previously accessible only through hydrothermal conversion—by pulsed laser deposition, and did so with a level of interfacial control that chemistry-based routes cannot provide. The strain-retained, stoichiometric STO shell boosts the photocurrent density of $TiO_2$ nanorod arrays to 1.9 mA $cm^{-2}$ at 1.23 V vs. RHE, exceeding the hydrothermal benchmark while retaining the one-dimensional transport geometry. Three independent lines of evidence converge on a single microscopic picture: the distorted yet oxygen-stoichiometric shell lattice establishes strain-induced ferroelectricity (structural), and KPFM images a built-in field that scales with device performance (electrostatic). Because every link of the strain–polarization–field–separation chain is individually verified and quantitatively cross-validated, this work transforms ferroelectric photoanode design from empirical optimization into an atomically engineerable, predictive strategy—one that is directly transferable to visible-light absorbers and multilayer ferroelectric architectures.

**Experimental section**

*Fabrication of $TiO_2$ nanorod arrays:* FTO substrates were cleaned with deionized water and dried under $N_2$. The conductive side was identified using a multimeter. The growth solution was prepared by mixing 15 mL of concentrated HCl (36.5–38 %) with 15 mL deionized water, followed by slow addition of 0.35 mL titanium tetra-butoxide under stirring. The solution was transferred to a 100 mL Teflon-lined autoclave, and FTO substrates were placed at an angle with the conductive side facing down. Hydrothermal growth was carried out at 150 °C for 18 h. After natural cooling, the samples were rinsed with deionized water and ethanol, and then annealed in air at 550 °C for 3 h to crystallize the $TiO_2$ nanorod arrays.

*PLD deposition of $SrTiO_3$:* $SrTiO_3$ was deposited onto the $TiO_2$ nanorod arrays by PLD using a KrF excimer laser (λ = 248 nm). The substrate temperature was maintained at 645 °C, the $O_2$ background pressure was 1.6 Pa, the laser energy was 55 mJ, and the repetition rate was 4 Hz. Deposition times of 7.5, 15, and 20 mins were used to vary the $SrTiO_3$ shell thickness.

*Post-deposition annealing:* For the annealing comparison, selected samples were heated in a muffle furnace at 500 °C for 1 h in air.

*Characterization:* The surface morphology and microstructure of the photoanodes were examined using a field-emission scanning electron microscopy (SEM, JSM-7800F, JEOL, Japan). High-resolution TEM (HRTEM) and energy-dispersive X-ray spectroscopy (EDX) for elemental analysis and mapping were performed using a JEOL JEM-2100 microscope to investigate the interfacial structure, lattice fringes, and crystallinity of the samples. X-ray photoelectron spectroscopy (XPS) measurements were conducted on a Thermo Scientific ESCALAB 250Xi spectrometer to determine the surface chemical composition and elemental oxidation states. UV–vis absorption spectra were recorded over the wavelength range of 300–800 nm using an Agilent Cary 5000 UV spectrophotometer, with $BaSO_4$ as the reference.

*Poling treatment:* Poling was performed at room temperature for 10 mins using a sandwich configuration, with FTO-coated glass as the top electrode and the sample NRs supported on an FTO-coated glass substrate as the bottom electrode. A poling voltage of ± 10 V was applied. For positive poling, a positive potential was applied to the bottom electrode.

*Photoelectrochemical measurements:* Mott–Schottky analysis was carried out in a three-electrode configuration in 1 M KOH at a fixed frequency of 1 kHz in the dark, using the photoanode as the working electrode, Ag/AgCl as the reference electrode, and a Pt plate as the counter electrode. J-V measurements were performed in a standard three-electrode configuration using a Pt wire as the counter electrode, Ag/AgCl (3 M NaCl) as the reference electrode, and an FTO-based sample as the working electrode.

Only the nanorod-loaded area was immersed in 1 M KOH electrolyte, and the exposed FTO contact was kept outside the solution. Illumination was provided by a 150 W Xe solar simulator equipped with an AM 1.5G filter, calibrated to 100 mW/cm² (1 sun) using a reference silicon solar cell.

*Kelvin probe force microscopy:* Amplitude-modulated Kelvin probe force microscopy (AM-KPFM) characterizations were carried out on an Asylum Research Cypher S AFM (Oxford Instruments) in single-pass mode under ambient atmosphere. Topography and contact potential difference (CPD) signals were acquired synchronously using a Ti/Ir-coated conductive probe (ASYELEC-01-R2, nominal k = 2.8 N/m, $f_0$ = 75 kHz, tip radius = 25 ± 10 nm). Images were captured over a 20 × 20 μm area with a scan rate of 1 Hz.

**Acknowledgments:**

This research was supported by the National Natural Science Foundation of China (No.62574133), Shanghai Pujiang Program (No. 24PJA083), and Natural Science Foundation of Zhejiang Province (LZ25F040003).

**Competing interests:**

The authors declare no conflict of interest.

## References

[1] A. Fujishima, K. Honda, *Nature* **1972**, *238*, 37-38. https://doi.org/10.1038/238037a0.

[2] M. Grätzel, *Nature* **2001**, *414*, 338-344. https://doi.org/10.1038/35104607.

[3] M. G. Walter, E. L. Warren, J. R. McKone, S. W. Boettcher, Q. Mi, E. A. Santori,

N. S. Lewis, *Chem.Rev.* **2010**, *110*, 6446-6473. https://doi.org/10.1021/cr1002326.

[4] K. Sivula, R. van de Krol, *Nat. Rev. Mater.* **2016**, *1*, 15010. https://doi.org/10.1038/natrevmats.2015.10.

[5] Y. Yu, Z. Zhang, X. Yin, A. Kvit, Q. Liao, Z. Kang, X. Yan, Y. Zhang, X. Wang, *Nat . Energy* **2017**, *2*, 17045. https://doi.org/10.1038/nenergy.2017.45.

[6] S. Corby, R. R. Rao, L. Steier, J. R. Durrant, *Nat. Rev. Mater.* **2021**, *6*, 1136-1155. https://doi.org/10.1038/s41578-021-00343-7.

[7] J. H. Kim, D. Hansora, P. Sharma, J.-W. Jang, J. S. Lee, *Chem. Soc. Rev.* **2019**, *48*, 1908-1971. https://doi.org/10.1039/c8cs00699g.

[8] X. Wang, B. Wang, L. Mayrhofer, X. Meng, H. Shen, J. Chu, *Materials Science and Engineering: R: Reports* **2025**, *163*, 100918. https://doi.org/https://doi.org/10.1016/j.mser.2024.100918.

[9] B. Wang, C. Ji, H. Shen, L. Mayrhofer, X. Wang, J. Chu, *Renewable and Sustainable Energy Reviews* **2026**, *242*, 117311. https://doi.org/https://doi.org/10.1016/j.rser.2026.117311.

[10] L. Liu, X. Chen, *Chem.Rev.* **2014**, *114*, 9890-9918. https://doi.org/10.1021/cr400624r.

[11] X. Liu, G. Zhu, X. Wang, X. Yuan, T. Lin, F. Huang, *Adv. Energy Mater.* **2016**, *6*, 1600452. https://doi.org/https://doi.org/10.1002/aenm.201600452.

[12] X. Wang, L. Mayrhofer, M. Hoefer, S. Estrade, L. Lopez-Conesa, H. Zhou, Y. Lin, F. Peiró, Z. Fan, H. Shen, L. Schaefer, M. Moseler, G. Braeuer, A. Waag,

*Adv. Energy Mater.* **2019**, *9*, 1900725. https://doi.org/https://doi.org/10.1002/aenm.201900725.

[13] W. Zhang, H. He, H. Li, L. Duan, L. Zu, Y. Zhai, W. Li, L. Wang, H. Fu, D. Zhao, *Adv. Energy Mater.* **2021**, *11*, 2003303. https://doi.org/https://doi.org/10.1002/aenm.202003303.

[14] X. Wang, L. Mayrhofer, M. Keunecke, S. Estrade, L. Lopez-Conesa, M. Moseler, A. Waag, L. Schaefer, W. Shi, X. Meng, J. Chu, Z. Fan, H. Shen, *Small* **2022**, *18*, 2204136. https://doi.org/https://doi.org/10.1002/smll.202204136.

[15] F. Wu, Y. Yu, H. Yang, L. N. German, Z. Li, J. Chen, W. Yang, L. Huang, W. Shi, L. Wang, X. Wang, *Adv. Mater.* **2017**, *29*, 1701432. https://doi.org/https://doi.org/10.1002/adma.201701432.

[16] G. Jia, F. Sun, T. Zhou, Y. Wang, X. Cui, Z. Guo, F. Fan, J. C. Yu, *Nat. Commun.* **2024**, *15*, 4746. https://doi.org/10.1038/s41467-024-49168-2.

[17] H. Yu, F. Chen, X. Li, H. Huang, Q. Zhang, S. Su, K. Wang, E. Mao, B. Mei, G. Mul, T. Ma, Y. Zhang, *Nat. Commun.* **2021**, *12*, 4594. https://doi.org/10.1038/s41467-021-24882-3.

[18] P. R. Willmott, J. R. Huber, *Reviews of Modern Physics* **2000**, *72*, 315-328. https://doi.org/10.1103/RevModPhys.72.315.

[19] H. M. Christen, G. Eres, *Journal of Physics: Condensed Matter* **2008**, *20*, 264005. https://doi.org/10.1088/0953-8984/20/26/264005.

[20] P. R. Willmott, *Progress in Surface Science* **2004**, *76*, 163-217. https://doi.org/https://doi.org/10.1016/j.progsurf.2004.06.001.

[21] Y. Kim, Y. Choi, S. A. Lee, W. S. Choi, K. T. Kang, *Current Applied Physics* **2024**, *68*, 113-130. https://doi.org/https://doi.org/10.1016/j.cap.2024.09.001.

[22] X. Gao, X. Ding, X. Ren, K. H. L. Zhang, D. Bui, *Appl. Surf. Sci.* **2025**, *709*, 163802. https://doi.org/https://doi.org/10.1016/j.apsusc.2025.163802.

[23] H.-C. Ho, M. Smiljanić, Z. Jovanović, M. Čekada, J. Kovač, G. Koster, J. i. Hlinka, N. Hodnik, M. Spreitzer, *ACS Appl. Mater. Interfaces* **2023**, *15*, 44482-44492. https://doi.org/10.1021/acsami.3c07747.

[24] X. Feng, K. Shankar, O. K. Varghese, M. Paulose, T. J. Latempa, C. A. Grimes, *Nano Lett.* **2008**, *8*, 3781-3786. https://doi.org/10.1021/nl802096a.

[25] R. P. Vasquez, *Surface Science Spectra* **1992**, *1*, 129-135. https://doi.org/10.1116/1.1247683.

[26] R. J. Cottier, N. A. Steinle, D. A. Currie, N. Theodoropoulou, *Appl. Phys. Lett.* **2015**, *107*. https://doi.org/10.1063/1.4936608.

[27] J. H. Haeni, P. Irvin, W. Chang, R. Uecker, P. Reiche, Y. L. Li, S. Choudhury, W. Tian, M. E. Hawley, B. Craigo, A. K. Tagantsev, X. Q. Pan, S. K. Streiffer, L. Q. Chen, S. W. Kirchoefer, J. Levy, D. G. Schlom, *Nature* **2004**, *430*, 758-761. https://doi.org/10.1038/nature02773.

[28] S. Assavachin, C. Xiao, H. Vahdatvasei, T. Mamani, D. Donadio, F. E. Osterloh, *EES Catalysis* **2025**, *3*, 1315-1326. https://doi.org/10.1039/d5ey00266d.

[29] W. Melitz, J. Shen, A. C. Kummel, S. Lee, *Surf. Sci. Rep.* **2011**, *66*, 1-27. https://doi.org/https://doi.org/10.1016/j.surfrep.2010.10.001.

[30] A. Zahmatkeshsaredorahi, R. Millan-Solsona, D. S. Jakob, L. Collins, X. G. Xu,

*Nature Reviews Methods Primers* **2025**, *5*, 53. https://doi.org/10.1038/s43586-025-00424-9.

[31] J. Xie, C. Guo, P. Yang, X. Wang, D. Liu, C. M. Li, *Nano Energy* **2017**, *31*, 28-36. https://doi.org/https://doi.org/10.1016/j.nanoen.2016.10.048.

## Figure captions

**Fig. 1.** (a) Fabrication procedure of the $TiO_2$-$SrTiO_3$ core–shell photoanode. (b) Experimental setup of the PLD. (c–d) SEM images of $TiO_2$ and $TiO_2$–$SrTiO_3$. (e) XPS O 1s spectra of $TiO_2$ and $TiO_2$–$SrTiO_3$. (f) EDX mapping of $TiO_2$–$SrTiO_3$. (g) HRTEM image of $TiO_2$–$SrTiO_3$. FFT images of the $TiO_2$ core and $SrTiO_3$ shell are shown in the inset.

**Fig. 2.** (a) UV-Vis spectra of $TiO_2$ and $TiO_2$-$SrTiO_3$ with 7, 15, 20 mins. (b) The corresponding Tauc plots of $TiO_2$ and $TiO_2$-$SrTiO_3$ with 7, 15, 20 mins.

**Fig. 3.** (a) Mott-Schottky plots of $TiO_2$ and $TiO_2$-$SrTiO_3$ with 7, 15, 20 mins. Enlarged M-S plots of $TiO_2$-$SrTiO_3$ with 7, 15, 20 mins are shown in the inset. J-V curves of $TiO_2$ and $TiO_2$-$SrTiO_3$ with 7, 15, 20 mins at (b) no, (c) +p, and (d) -p poling.

**Fig. 4.** KPFM of $TiO_2$ in (a) dark, (b) under light, and (c,d) the corresponding histograms. KPFM of $TiO_2$-$SrTiO_3$ with 15 mins in (e) dark, (f) under light, and (g,h) the corresponding histograms.

**Fig. 5.** Band diagrams of $TiO_2$-$SrTiO_3$ for (a) +p, (b) no, and (c) -p poling.

**Table 1.** Donor densities and onset potentials of $TiO_2$ and $TiO_2$-$SrTiO_3$ with 7, 15, 20 mins.

## Fig. 1

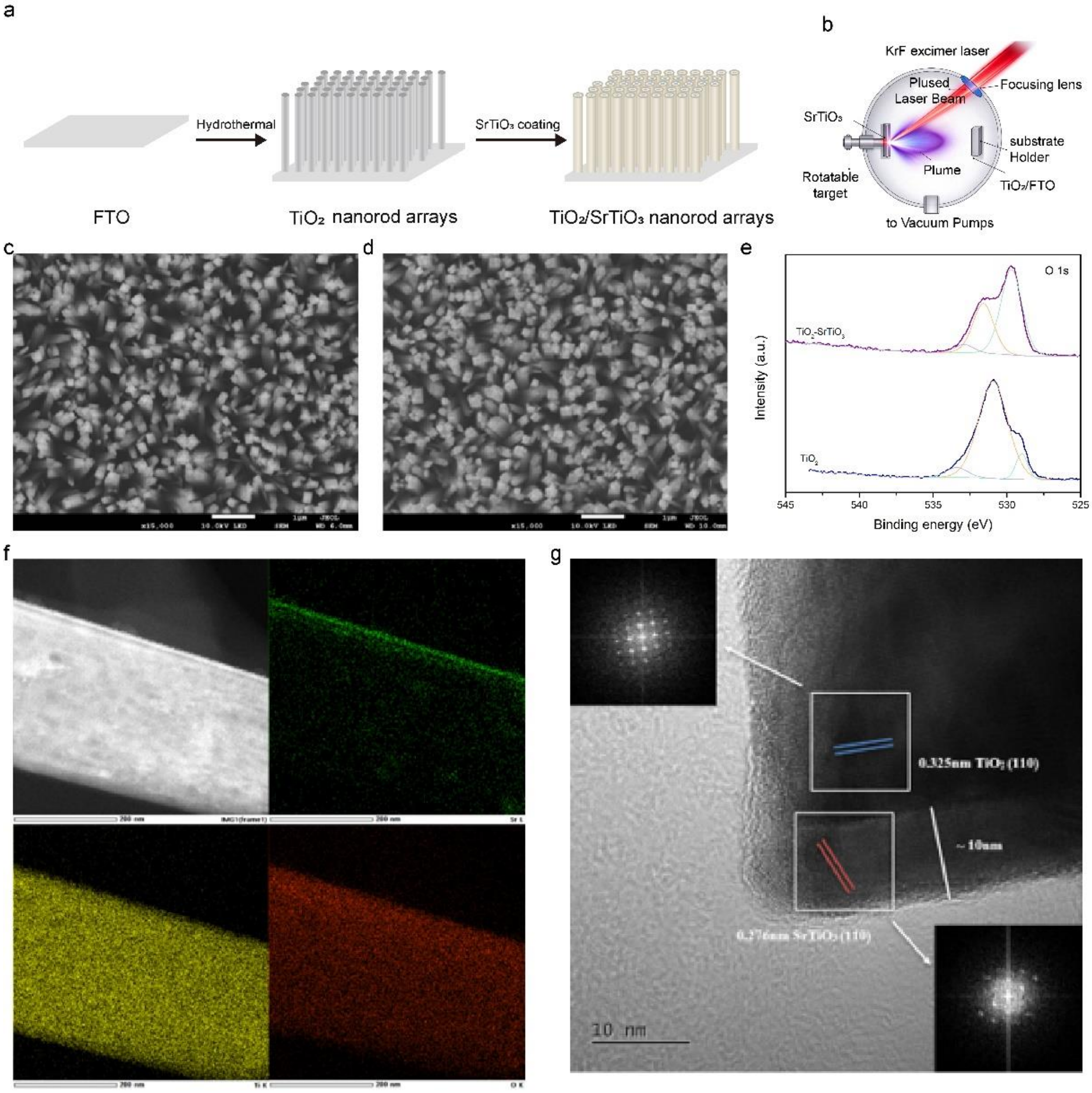
a
Hydrothermal
SrTiO₃ coating
FTO
TiO₂ nanorod arrays
TiO₂/SrTiO₃ nanorod arrays
b
KrF excimer laser
Plused Laser Beam
Focusing lens
SrTiO₃
substrate Holder
Plume
Rotatable target
TiO₂/FTO
to Vacuum Pumps
c
d
e
O 1s
TiO₂-SrTiO₃
TiO₂
Intensity (a.u.)
Binding energy (eV)
545
540
535
530
525
f
g
0.325nm TiO₂ (110)
~10nm
0.276nm SrTiO₃ (110)
10 nm

**Fig. 2**

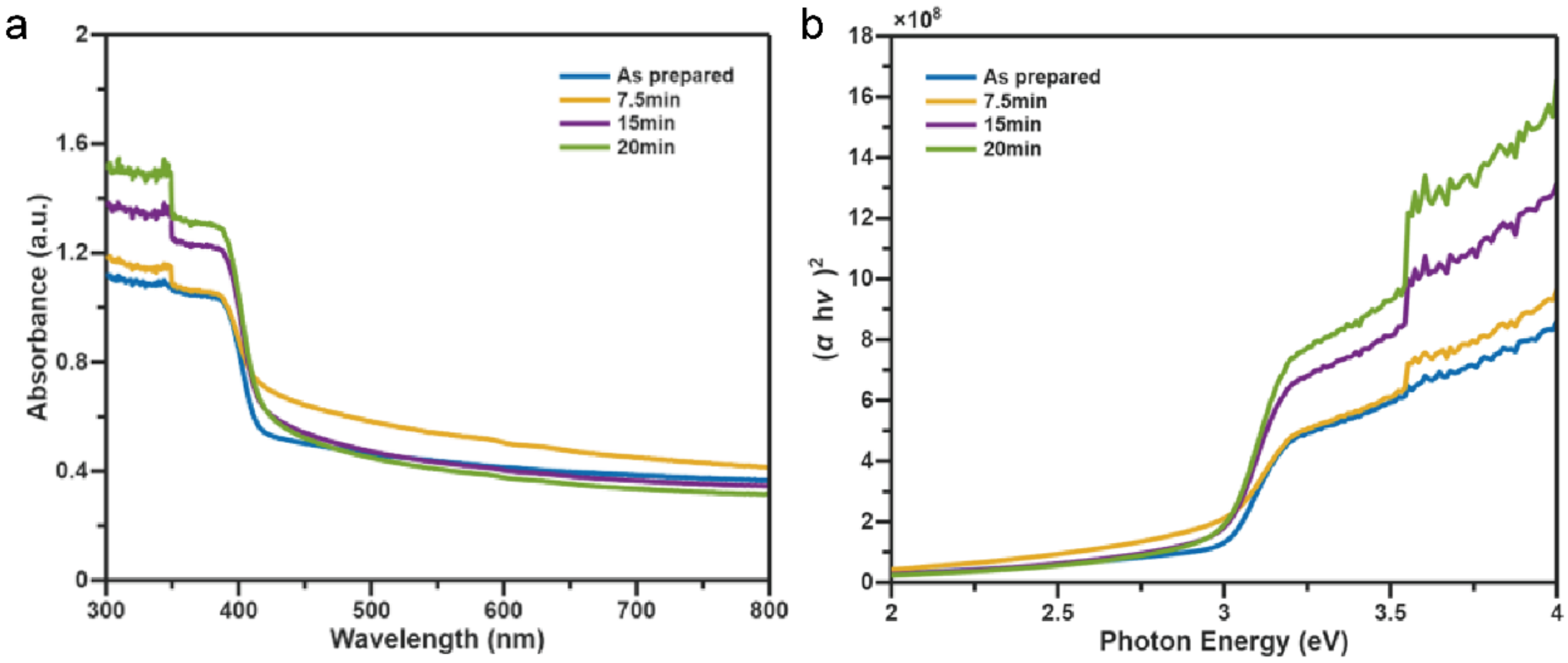

a
As prepared
7.5min
15min
20min
Absorbance (a.u.)
Wavelength (nm)
b
×10^8
As prepared
7.5min
15min
20min
(α hν)^2
Photon Energy (eV)

**Fig. 3**

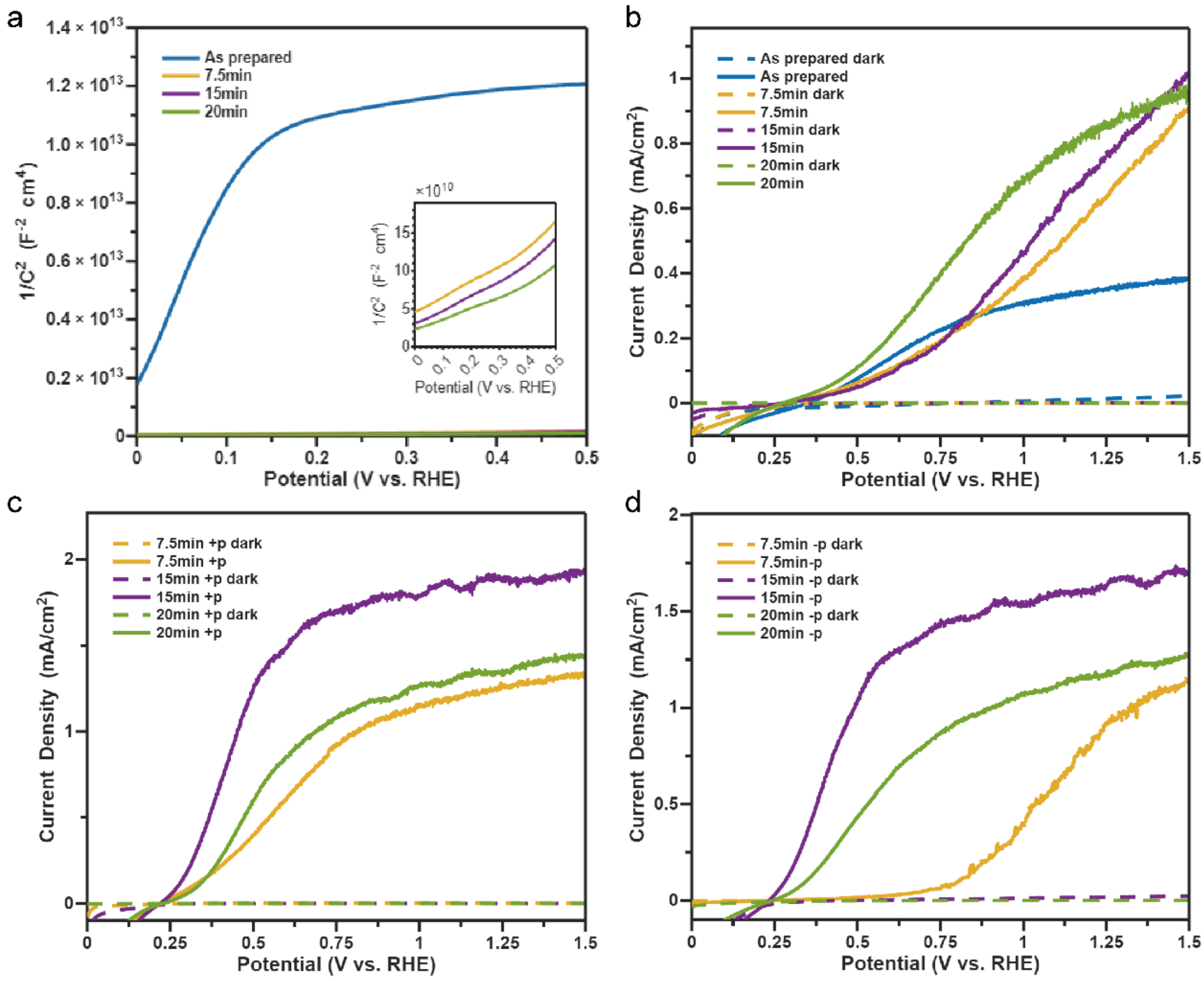
a
1.4 × 10^13
1.2 × 10^13
1.0 × 10^13
0.8 × 10^13
0.6 × 10^13
0.4 × 10^13
0.2 × 10^13
0
1/C² (F⁻² cm⁴)
As prepared
7.5min
15min
20min
×10^10
15
10
5
0
0 0.1 0.2 0.3 0.4 0.5
Potential (V vs. RHE)
0 0.1 0.2 0.3 0.4 0.5
Potential (V vs. RHE)
b
As prepared dark
As prepared
7.5min dark
7.5min
15min dark
15min
20min dark
20min
Current Density (mA/cm²)
1
0.8
0.6
0.4
0.2
0
0 0.25 0.5 0.75 1 1.25 1.5
Potential (V vs. RHE)
c
7.5min +p dark
7.5min +p
15min +p dark
15min +p
20min +p dark
20min +p
Current Density (mA/cm²)
2
1
0
0 0.25 0.5 0.75 1 1.25 1.5
Potential (V vs. RHE)
d
7.5min -p dark
7.5min-p
15min -p dark
15min -p
20min -p dark
20min -p
Current Density (mA/cm²)
2
1.5
1
0.5
0
0 0.25 0.5 0.75 1 1.25 1.5
Potential (V vs. RHE)

**Fig. 4**

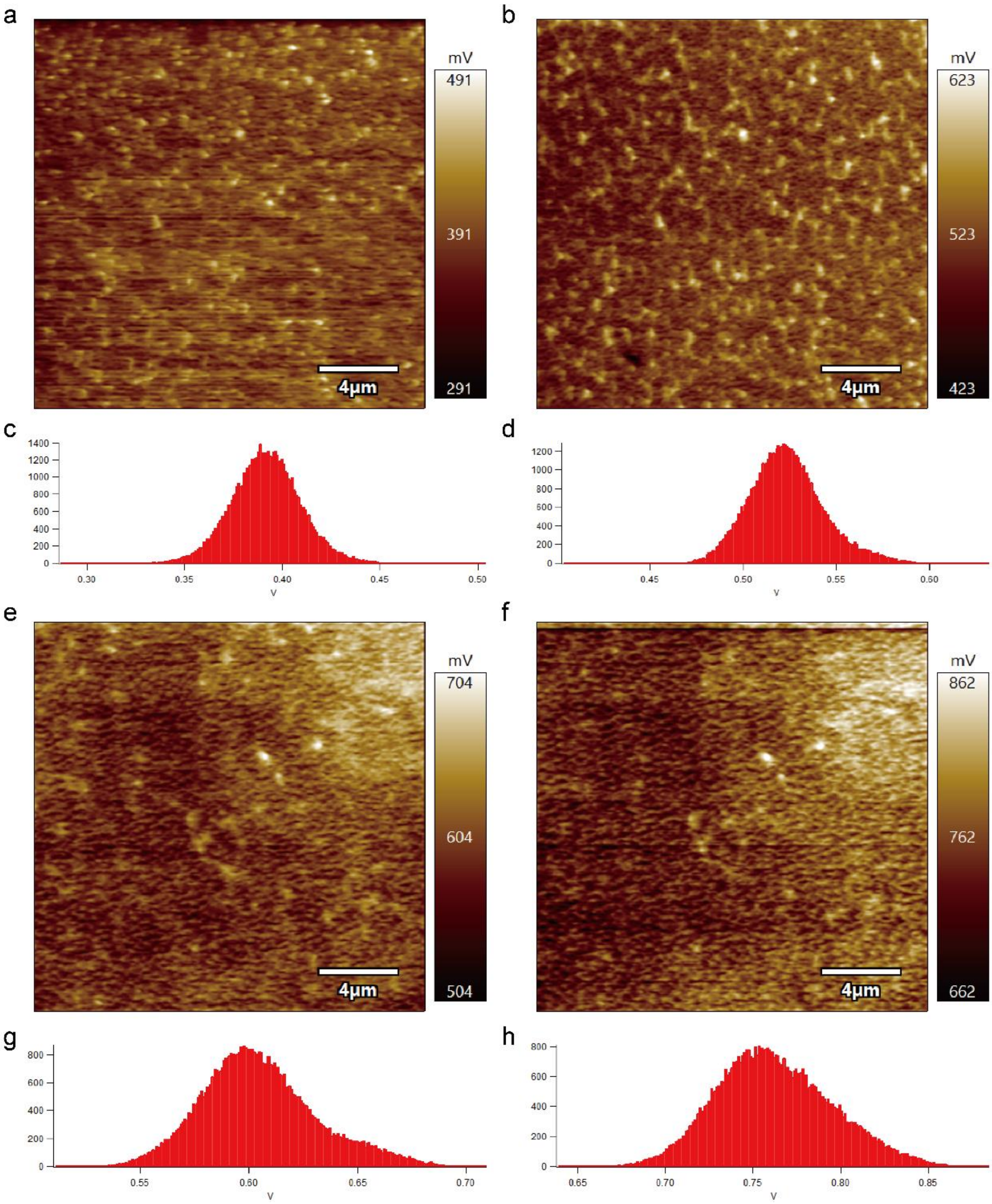

a
b
mV
491
391
291
4μm
mV
623
523
423
4μm
c
d
V
V
e
f
mV
704
604
504
4μm
mV
862
762
662
4μm
g
h
V
V

**Fig. 5**

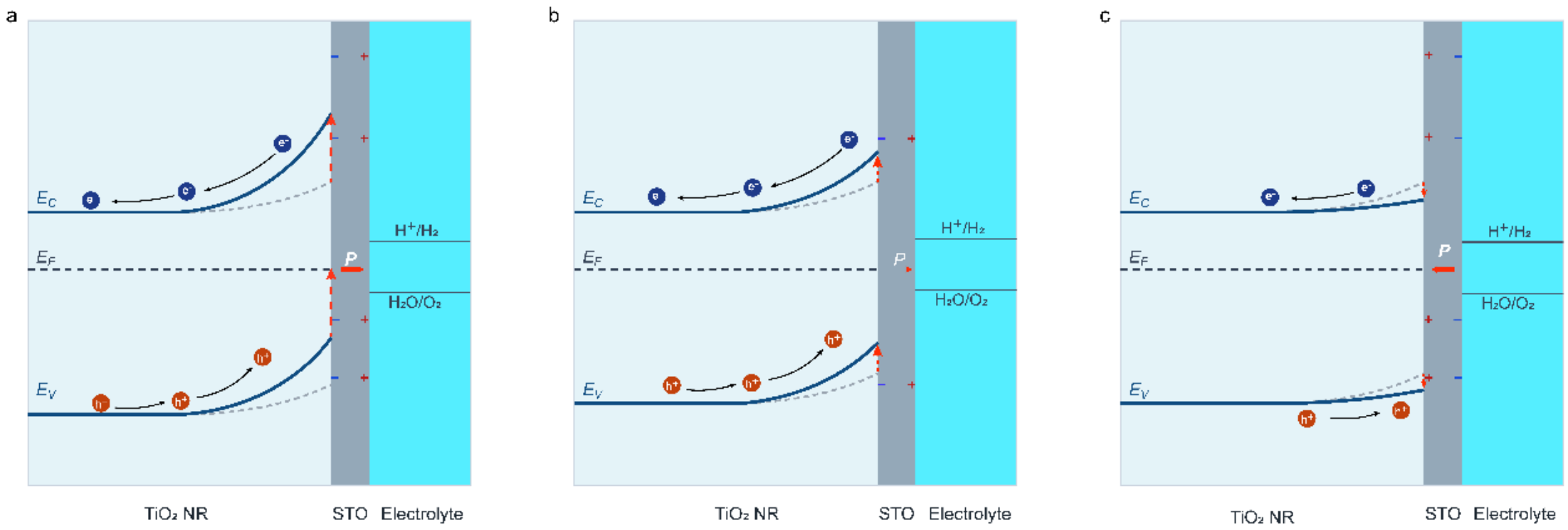

**Table 1**

| Sample | $N_d$ ($cm^{-3}$) | Onset potential (V) |
| --- | --- | --- |
| $TiO_2$ | $2.496 \times 10^{16}$ | −0.384 |
| $TiO_2$-$SrTiO_3$, 7.5 mins | $9.068 \times 10^{19}$ | −0.186 |
| $TiO_2$-$SrTiO_3$, 15 mins | $1.095 \times 10^{19}$ | −0.119 |
| $TiO_2$-$SrTiO_3$, 20 mins | $1.449\times 10^{19}$ | −0.119 |